\documentclass[draft]{agujournal2019}
\usepackage{url} 
\usepackage{lineno}
\usepackage[inline]{trackchanges} 
\usepackage{soul}
\usepackage{amsmath,amssymb}  
\drafttrue
\journalname{JGR: Planets}

\begin{document}

%
%


\title{High-Latitude Zonal Jets in the Martian Upper Atmosphere Driven by Non-Orographic Gravity Waves}
%
%




\authors{Jiandong Liu\affil{1,2,*}, François Forget\affil{1}, Ehouarn Millour\affil{1}, Francisco González Galindo\affil{3}, Jean-Yves Chaufray\affil{4}}


\affiliation{1}{LMD/IPSL, Sorbonne Université, ENS, Université PSL, École Polytechnique, Institut Polytechnique de Paris, CNRS, Paris, France.}
\affiliation{2}{Laboratoire de Physique et Chimie de l’Environnement et de l’Espace, Université d’Orléans, CNRS, Orléans, France.}
\affiliation{3}{Instituto de Astrofisica de Andalucia, Granada, Spain.}
\affiliation{4}{LATMOS/IPSL, Sorbonne Université, UVSQ, CNRS, Paris, France}




\correspondingauthor{Jiandong Liu}{Jiandong.liu@lmd.ipsl.fr}




\begin{keypoints}
\item Simulations show that non-orographic gravity waves modulate the winds throughout the Martian upper atmosphere.
\item High-latitude zonal jets are strongly regulated in the hemisphere where the descending branches of the Hadley cell are located.
\item This regulation aligns with wind observations from the Neutral Gas and Ion Mass Spectrometer.
\end{keypoints}

%
%

%
%


\begin{abstract}
We investigate thermosphere responses to non-orographic gravity waves (GWs) using wind measurements from the Neutral Gas and Ion Mass Spectrometer onboard the Mars Atmosphere and Volatile EvolutioN mission, alongside simulations from the Mars Planetary Climate Model. We focus on zonal jets in high-latitude regions of the upper atmosphere. Jet acceleration and deceleration ($\pm$280 m s$^{-1}$) arise from momentum divergence ($\pm$ 1300 m s$^{-1}$ sol$^{-1}$) driven by wave saturation and wind filtering. Simulations and observations indicate that GWs modulate these jets in the hemisphere associated with the descending branches of the Hadley Cell, due to the absence of wave critical layers in the middle atmosphere. Interactions between GWs and the mean flow can shape the circulation and dynamics of the upper atmosphere of Mars.
\end{abstract}

\section*{Plain Language Summary}
This study examines non-orographic gravity waves in the atmosphere of Mars. These waves are generated by rising air in the lower atmosphere, such as from convection, and by large-scale motions such as jet streams and weather fronts. Using the Mars Planetary Climate Model and wind measurements from the Neutral Gas and Ion Mass Spectrometer onboard the Mars Atmosphere and Volatile EvolutioN mission, we studied how these waves affect high-latitude winds above 80 km. We found that they modulate strong east–west winds, or zonal jets, particularly in the hemisphere where air is descending within the Martian Hadley Cell—a seasonally varying global circulation pattern in which air rises in the warmer hemisphere, flows across the equator at high altitude, sinks in the colder hemisphere, and then returns near the surface. Understanding these effects improves our knowledge of atmospheric circulation and gas transport on Mars.

%
%

%


%
%
%
%

\section{Introduction}
The upper atmospheric dynamics of Mars governs gas escapes that are key to understanding the evolution of planet habitability \cite{chaffin2017elevated,benna2019global,yiugit2023coupling}. 
The upper atmospheric winds, predominantly driven by solar energy inputs, undergo a diurnal cycle \cite{gonzalez2010thermal,benna2019global,roeten2022impacts} while exhibiting high temporal variability \cite{miyoshi2014global,liu2023surface}. Additionally, non-orographic gravity waves (GWs)
influence the rotation and temporal variability of the diurnal wind vector by transporting momentum and energy from below \cite{medvedev2011influence,yiugit2008parameterization, miyoshi2014global,kuroda2015global,kuroda2016global,kuroda2020gravity,roeten2022impacts, liu2023surface, yiugit2023coupling}. However, the thermospheric wind's responses to non-orographic GWs are poorly understood due to the scarcity of observations \cite{jurado2026cross}.

Simulations conducted by General Circulation Models (GCMs) show a general effect of waves-induced wind damping in the upper atmosphere, which results in thermospheric cooling \cite{medvedev2011influence, medvedev2013general, roeten2022impacts, roeten2022maven, liu2023surface}. Non-orographic GWs dissipate momentum and energy substantially in the Martian high-latitude regions of the middle atmosphere, as revealed by GW-resolved simulations \cite{kuroda2015global,kuroda2016global,kuroda2020gravity, kling2025impact}, triggering dramatic zonal wind variations ($|U|\geq250$ m s$^{-1}$) in the upper atmospheric polar region \cite{roeten2022impacts,liu2023surface,liu2025diurnal,liu2025stochastic}. Comparisons between simulations and observations have not yet been fully carried out in the upper polar regions to confirm the effect. 

Models with GW parameterization can capture the magnitude of the mean thermospheric wind speed ($|U|\in$ [0,350) and $|V|\in$ [0,60] m s$^{-1}$, depending on the season) derived from the observations \cite{benna2019global,roeten2022maven} of Neutral Gas and Ion Mass Spectrometer onboard the Mars Atmosphere and Volatile EvolutioN mission (NGIMS/MAVEN) . However, these comparisons are conducted without considering the instantaneous variations in the wind vectors.
Moreover, other wave effects on the winds have not yet been adequately investigated, especially their contribution to the acceleration of upper-atmospheric jets. Here, we refer to the accelerating influence as a strengthening of wind velocity by GWs. This effect has been captured by GCMs \cite{miyoshi2014global,liu2023surface,liu2025diurnal,liu2025stochastic}, but has not been compared with the winds derived from NGIMS in the upper atmosphere. 

Intense middle and upper atmospheric wind responses to non-orographic GWs have been simulated by several Mars GCMs \cite{miyoshi2008gravity, miyoshi2014global, medvedev2013general, roeten2022impacts, liu2023surface}. \citeA{medvedev2011influence} reported GW-induced damping of mean flows below 160 km, which hits its maximum during global dust storms \cite{medvedev2013general}. \citeA{roeten2022impacts} reproduced the damping with a whole atmosphere GCM using the same GW scheme as in \citeA{yiugit2008parameterization} and \citeA{medvedev2013general}. As a result, the simulated mean winds are compatible with the averaged NGIMS' observations in the upper atmosphere \cite{roeten2022impacts,roeten2022maven}. The damping effect is also captured by the wave-resolved model in the middle atmosphere \cite{miyoshi2008gravity, miyoshi2014global, kuroda2015global,kuroda2016global,kuroda2020gravity,kling2025impact}.
The effect is so large that the waves tend to close the simulated mesospheric jets \cite{kuroda2015global,kuroda2020gravity}. \citeA{liu2023surface} shows that the damping or acceleration of the winds by GWs depends on the season and local solar time \cite{miyoshi2014global}. Simulations show that wind vectors exhibit diurnal rotation with strong instantaneous fluctuations, synchronous with the diurnal thermal tide \cite{gonzalez2010thermal, benna2019global, liu2023surface}.
These simulations revealed the wind climatology in the upper atmosphere. However, the GWs-jets-circulations coupling remains unclear.

High wind variability has been captured by measurements throughout the middle and upper atmosphere, despite the limited spatiotemporal coverage of the observations \cite{jurado2026cross}.
 Zonal wind spikes in the middle atmosphere have been observed by retrievals from the radio telescope during Martian global dust storms \cite{lellouch1991first,cavalie2008vertical,moreno2009wind,miyamoto2021intense}. Orbit-to-orbit wind variability is also high ($\pm$150 m s$^{-1}$) as measured by NGIMS in the upper atmosphere \cite{roeten2019maven,benna2019global,roeten2022impacts,roeten2022maven}. 
 NGIMS-derived winds have yielded meaningful modeling constraints, including estimates of wind magnitude \cite{roeten2022maven}. The high variability in the instantaneous observations is not reproduced by the model \cite{benna2019global,roeten2022impacts}. This behavior manifests itself in two ways: strong orbit-to-orbit variability ($\sim$ 150 m s$^{-1}$ in the zonal direction and $\sim$ 40 m s$^{-1}$ in the meridional direction) at fixed locations and local solar time, and abrupt changes (spikes) in wind direction and velocity ($|U|_{max}>400$ and $|V|_{max}>80$ m s$^{-1}$) within short time intervals (in minutes).

Consequently, we must focus on the regions that are significantly impacted by non-orographic GWs to investigate the waves-winds coupling. Saturated non-orographic GWs can release large amounts of momentum in the high-latitudes of the upper atmosphere \cite{benna2019global,roeten2019maven,roeten2022impacts,roeten2022maven, liu2023surface}. Zonal winds are substantially altered in these regions \cite{liu2023surface,kling2025impact}. To compare the simulation with the observation meaningfully, the dominant wind diurnal cycle driven by solar forcing and the local wind variability triggered by other factors may be considered separately. The comparison shall take place in areas where the wind behaviors are strongly modulated by the non-orographic GWs.

In this paper, we investigated the behaviors of zonal jets in the upper atmospheric high-latitudes under the impacts of non-orographic GWs. The simulations are conducted 
with the Mars Planetary Climate Model (Mars PCM) and a non-orographic GW scheme designed by \citeA{liu2023surface}. The measurements are the instantaneous winds sampled by NGIMS \cite{elrod2014maven}. 
Section 2 describes the non-orographic GW parameterization in the Mars PCM and the observational geometry of NGIMS. Section 3 presents the simulated zonal jets. Section 4 shows the PCM-NGIMS comparisons. Section 5 discusses the wave-zonal wind interactions based on selected cases. The summary is given in section 6.

\section{Model Settings and NGIMS Observational Geometry }
\subsection{Wave-Mean Flow Interactions}
The Mars Planetary Climate Model, formerly known as the Laboratoire de Météorologie Dynamique Mars General Climate Model (LMD-MGCM), \cite{forget1999improved}  applies a stochastic non-orographic GW scheme \cite{liu2023surface}. The scheme represents non-orographic GWs using wave ensembles, in which each ensemble includes eight monochromatic waves with stochastic horizontal wavenumbers, phase speeds, and momentum flux \cite{lott2012stochastic,lott2013stochastic}. It has been tuned along with the model's dust-water cycle. The simulations are compatible with the temperatures measured by the Mars Climate Sounder (MCS) and the abundances sampled by the NGIMS \cite{liu2023surface,liu2025diurnal,liu2025stochastic}.

The scheme computes the vertical evolution of the waves by evaluating the equivalent vertical component of Eliassen-Palm flux (EP-flux) that represents the pseudo-momentum (vertical flux of zonal momentum). A pseudo-momentum, $\vec{E}_j=\rho \overline{u'w'}$, on the order of $0\text{-}5\times10^{-4}\,\mathrm{kg\,m^{-1}\,s^{-2}}$ is initialized at an altitude of 5.8~km, with horizontal propagation restricted to the zonal direction only, that is, eastward or westward. It decays due to critical layers, kinematic viscosity, turbulence damping, and saturation once launched \cite{lott2012stochastic,liu2023surface}. The divergence of EP-flux $\frac{1}{\rho}\frac{d \vec{E}_j}{dz}$ is averaged between two adjacent time steps $t+\delta t$ and $t$, and is added to the mean flow $\vec{\bar{u}}$ using the first-order autoregressive process,
\begin{equation}
\Bigg ( \frac{\partial \vec{\bar{u}}}{\partial t} \Bigg )_{GW}^{t+\delta t} =\frac{\delta t }{\Delta t} \frac{1}{M} \sum_{j=1}^{M} \frac{1}{\rho} \frac{d \vec{E}_j}{dz} + \frac{\Delta t - \delta t}{\Delta t} \Bigg ( \frac{\partial \vec{\bar{u}}}{\partial t} \Bigg )_{GW}^{t}
\label{divergens}
\end{equation}
With an average life cycle of 1 sol ($\Delta t = 88{,}775$ s), the wave is resolved in the model using a physical time step of $\delta t = 900$ s. Each ensemble includes M = 8 monochromatic waves.

Wave saturation and critical layers are the dominant mechanisms that regulate wave-flow interactions.
The typical saturation altitudes of the simulated waves are located in the lower thermosphere \cite{medvedev2011influence,roeten2019maven,liu2023surface,liu2025diurnal,liu2025stochastic}. Some waves can even reach altitudes of 180 km during dusty seasons \cite{roeten2019maven,liu2023surface}. The upper atmospheric winds, therefore, experience significant perturbations due to the wave-flow interactions, as shown in equation (\ref{divergens}). 

In addition, a critical layer is the altitude at which a harmonic dissipates because its intrinsic frequency, $\Omega=\vec{k}\cdot(\vec{k}c/|\vec{k}|-\vec{U})$, vanishes \cite{lott2012stochastic, liu2023surface, liu2025diurnal}. Here, $\vec{k}=(k,l)$ denotes the horizontal wavenumber, $c$ the phase speed, and $\vec{U}=(U,V)$ the background wind velocity. We have set $c\sim |N(0,U_{z_r}^2)|$, following a folded Gaussian distribution with variance $\sigma^2= U_{z_r}^2(1-2/\pi)$ and mean $U_{z_r}\sqrt{2/\pi}$; $U_{z_r}$ the square of the zonal wind $U$ at the wave source altitude $z_r$ \cite{lott2012stochastic,lott2013stochastic}. $\Omega$ is the factor that controls the wind-filtering effect. How the GW's parameters are determined and validated with observations is described in \citeA{liu2023surface}. The filtering effect is detailed in \citeA{liu2025diurnal}. The turbulent mixing of the GW is described in \citeA{liu2025stochastic}.

\subsection{Observational Geometry of NGIMS/MAVEN}
The NGIMS conducts wind observations by changing its pointing direction within a small angle ($\pm 8^{\circ}$) along with its controllable mounted platform \cite{elrod2014maven, benna2019global}. The uncertainties associated with the along-track and cross-track wind components are estimated to be 20 m s$^{-1}$ and 6 m s$^{-1}$, respectively. These uncertainties are small compared to the magnitudes of the jets in Figure \ref{simujets}. The campaign occurs at altitudes of 150 to 250 km \cite{benna2019global}. Winds are sampled in both inbound (descending) and outbound (ascending) legs at a frequency of 1 Hz, with a general period of 600 s. 

\begin{figure}
\noindent\includegraphics[width=14cm,trim={0.5cm 0.5cm 0 0cm}]{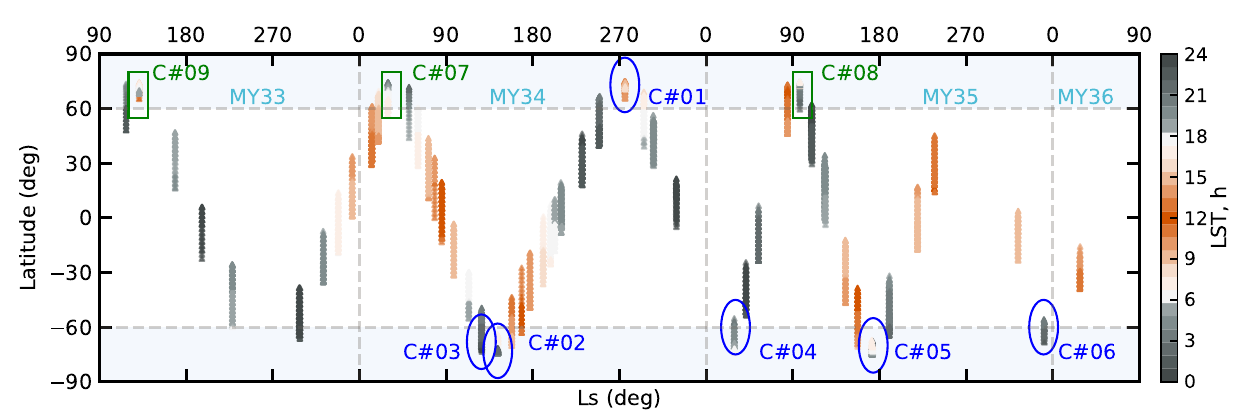}
\caption{Observational geometry of NGIMS/MAVEN during wind sampling from MY33 to MY36. Local Solar Time (LST) is color-coded along with latitude and solar longitude (Ls). High-latitude regions ($\geq$ $60^{\circ}$) are shaded in light gray. Samples \texttt{C\#}01 to \texttt{C\#}06 correspond to the descending branch of the Hadley Circulation (HC), whereas \texttt{C\#}7 to \texttt{C\#}9 correspond to the ascending branch. The dusty season occurs between Ls 135$^\circ$ and 270$^\circ$. Each sample includes 5-15 orbits.}
\label{obsgeo}
\end{figure}

Figure \ref{obsgeo} illustrates the observational geometry of NGIMS during MY33 to MY36. The Local Solar Times (LST) of the samples vary with latitude and Ls. Each of the samples at a given Ls includes 5-15 orbits.
Nine cases (\texttt{C\#}01 to \texttt{C\#}09) have been selected, in which the winds were sampled in the high-latitude region (latitude $>$ 60°; gray zones in Figure \ref{obsgeo}). \texttt{C\#}01 to \texttt{C\#}06 are located within the descending branch of the Hadley Cell (HC; Figure \ref{simujets}).
\texttt{C\#}07 to \texttt{C\#}09 are sampled within the ascending branch of the HC (Figure \ref{simujets}b, \ref{simujets}c, \ref{simujets}e). The HC is constructed by the zonal-mean meridional mass streamfunction in every 30$^\circ$ bins in Ls \cite{asumi2025climatology}, using the Mars PCM meridional winds $V$ and pressure \cite{liu2023surface}. This interpretation is also consistent with the established climatological Hadley circulation structure reported in previous Mars GCM studies \cite{liu2023surface,liu2025diurnal,asumi2025climatology}.
Unfortunately, only \texttt{C\#}01 occurred between 9-15 h, albeit during the polar night. The remaining cases were sampled near dawn and dusk terminators.

\section{Simulated Upper Atmospheric Zonal Jets}
We conducted simulations without (GWoff) and with (GWon) gravity waves from Mars Year (MY)29 to MY36 to investigate the behavior of the winds. Here, we refer the GWs to the non-orographic GWs since the orographic GW scheme is turned on by default in both cases. The zonal winds $U$ from the two cases are used to examine the climatology. The meridional wave drag is negligible, primarily because wave propagation is restricted to zonal directions at launch, so that the meridional momentum flux $\rho \overline{v'w'}$ is effectively zero. Consequently, little to no meridional wave drag is produced near the source altitudes \cite{liu2023surface,liu2025diurnal,liu2025stochastic}. The induced meridional winds generated by the zonal drags \cite{asumi2025climatology}, $V^*$, are negligible below the thermosphere but reach 40 to 80 m s$^{-1}$ in the upper atmosphere \cite{liu2025diurnal}. Here $V^*=\overline{V}-1/\rho (d\vec{E}/dz)|_U/ f$ \cite{liu2025diurnal, asumi2025climatology}, with $f$ denoting the Coriolis frequency. Zonal winds are directly affected by GWs and serve as a better indicator than meridional winds for detecting wave effects.

\begin{figure}
\noindent\includegraphics[width=14cm, trim={0.3cm 1cm 0 1.5cm}]{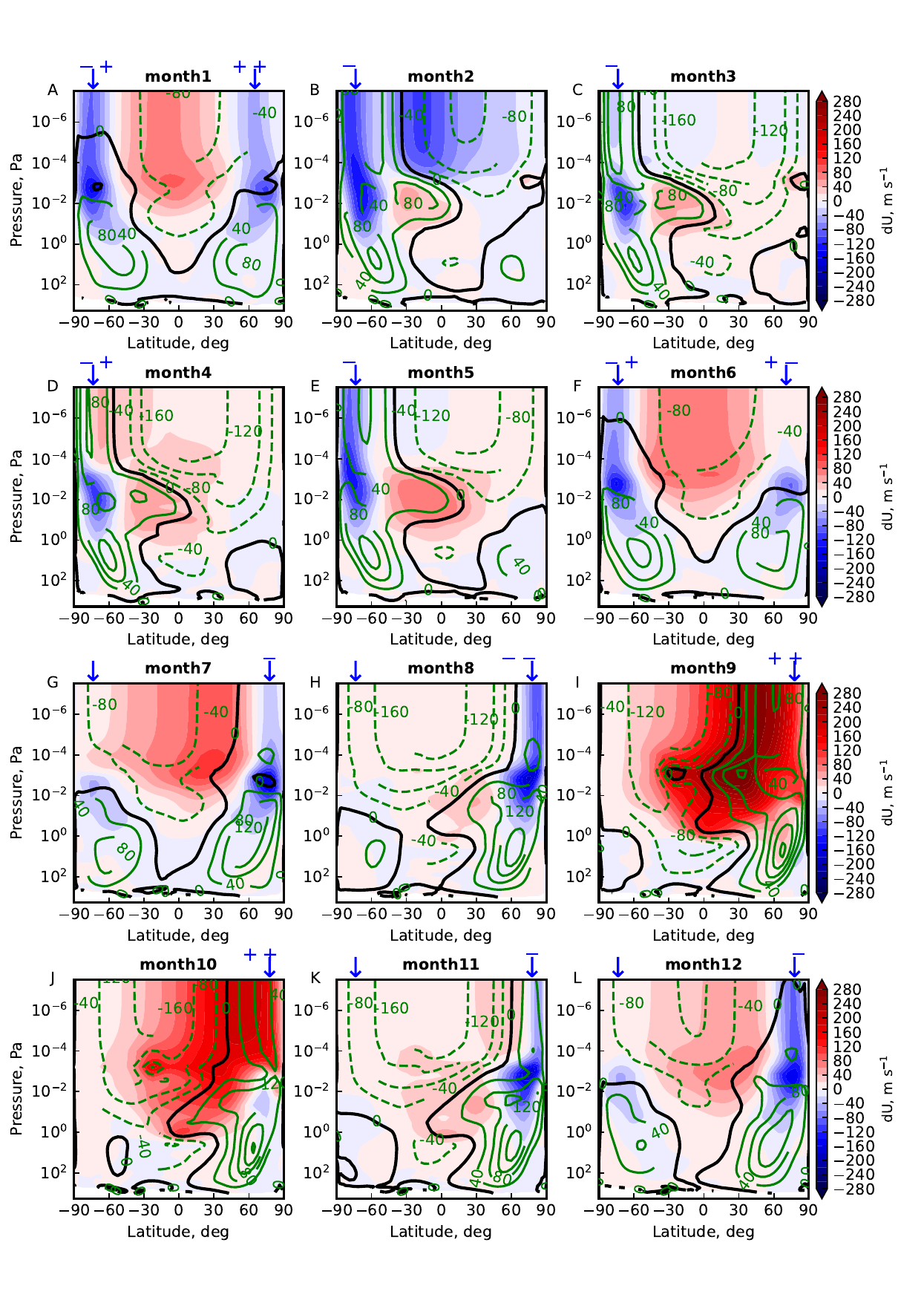}
\caption{Monthly-mean (30$^\circ$ bins in Ls) zonal-mean zonal winds ($U$) and wave-induced wind changes ($dU$) during MY29. Wave effects are defined as $dU = U_{\mathrm{GWon}} - U_{\mathrm{GWoff}}$ (m s$^{-1}$) and are color-coded. Green contours show $U_{\mathrm{GWon}}$ (m s$^{-1}$), and bold black contours denote $U_{\mathrm{GWon}} = 0$. Blue arrows indicate the zonal jets, coincident with the descending branch of the Hadley circulation; $-$ damping; $+$ accelerating; $--$ closing; $++$ opening; $-+$ or $+-$ denote the coexistence of damping and acceleration. Negative (positive) $dU$ indicates wave-induced damping (accelerating).}
\label{simujets}
\end{figure}

Figure \ref{simujets} presents the wave-induced effects on zonal winds, averaged both monthly and zonally. We overlay the simulated zonal winds ($U_{GWon}$, contours) onto the differences ($dU=U_{GWon}-U_{GWoff}$, color-coded map). Thus, the winds are accelerated by the GWs if $U_{GWon}$ and $dU$ have the same sign. Otherwise, the winds are damped.
In addition, simulations were conducted for Mars Year (MY) 29.
The magnitudes of the simulated jets exhibit interannual variability due to changes in Martian dust storms and solar forcing. However, they share the same climatology.

\subsection{Responses of the Zonal Jets to the GWs}
Zonal jets most strongly influenced by wave saturation are located above 80-100 km (10$^{-1}$-10$^{-2}$ Pa), a feature consistently reproduced in multiple simulations \cite{miyoshi2008gravity,miyoshi2014global,medvedev2011influence,medvedev2013general,roeten2022impacts,liu2023surface}. We classify the gravity wave effects on the zonal jets into four categories, as illustrated in Figure \ref{simujets} and summarized in Table \ref{table}.

The jets can be severely damped by the waves (U$\times$dU$<$0), in which their vertical extensions have been demolished (U $\approx$ 0 m s$^{-1}$). The effect is referred to as jets 'closing'. For example, the zonal jet in the northern polar region is closed by waves during Ls 150$^\circ$-180$^\circ$ (Figure \ref{simujets}f).
Alternatively, the upper zonal winds are strongly accelerated by the waves (U$\times$dU$>$0), leading to the formation of new jets. This phenomenon is referred to as jets 'opening'.
 For example, a new jet is opened in the northern polar region during Ls 240$^\circ$-270$^\circ$ (Figure \ref{simujets}i).
In addition, cases in which no existing jets are closed and no new jets are formed are classified as jets 'damping' (U$\times$dU$<$0) and jets 'accelerating' (U$\times$dU$>$0), respectively. The jet in the southern polar region is damped during Ls 30$^\circ$-60$^\circ$ (Figure \ref{simujets}b). In contrast, the jet in the northern polar region is accelerated during Ls 0$^\circ$-30$^\circ$ (Figure \ref{simujets}a).

\begin{table}
\caption{Categories of High-latitudes Jets' Behaviors Triggered by Non-orographic GW. Unit: m s$^{-1}$. Closing: old jet being destroyed; opening: new jet being created$^b$; damping: existed jet being attenuated; accelerating: existed jet being strengthening.}
\centering
\begin{tabular}{l| c c c c c c}
\hline
Categories  & U$_{GWon}$ & U$_{GWoff}$& dU & \textbf{sign}[$\phi$]$^a$  &Examples \\
\hline
   closing       & $\sim$0 & $>$100  & (-40,-280) & -& month8\\
   opening       & $>$100  & $\sim$0 & (40,280)   & +& month9;month10\\
   damping       & $>$0    & $>$40   & (-40,-280) & -& month2-3;month5;month7; month11-12\\
   accelerating  & $>$40   & $>$0    & (40,280)   & +& month1, month4;\\
\hline
\multicolumn{6}{l}{$^{a}$ $\phi=$dU$\times$U$_{GWon}$; dU=U$_{GWon}-$U$_{GWoff}$. $^{b}$ closing-opening happens due to -U$\rightarrow$0$\rightarrow$+U.}
\end{tabular}
\label{table}
\end{table}

Note that we describe these effects using 'monthly-averages' (every 30$^\circ$ in Ls, i.e., climatological means). In instantaneous snapshots, these effects can coexist or occur alternately. 

\subsection{Intensive Impacted Regions and Seasons}
Figure \ref{simujets} reveals that the effects mentioned in section 3.1 are predominantly vivid in the Martian descending branch(es) of the Hadley Cell (HC) located in high-latitudes. 
Useful articles on Martian overturning circulations by mass stream function may be found in \citeA{forget1999improved, richardson2002topographically,liu2023surface}. The fast westerlies, in the lower and middle atmosphere of the hemisphere where the downward HC branch is located (Figure \ref{simujets}), favor the propagation of GWs without encountering much 'wind filtering'. This leads the waves to diverge their momentum at higher altitudes in the branch(es). Drags calculated from both GW-resolved models and GCMs have confirmed this view \cite{medvedev2013general,liu2023surface,kling2025impact}.

The effects of 'opening' and 'closing' indicate that the jets are 'generated' or 'purged' by the waves. These effects are more pronounced during the dusty seasons than in the clean-sky seasons. Simulations have predicted a net increase in GW drags in the upper atmosphere during the dusty season \cite{roeten2019maven,liu2023surface,kling2025impact}, due to favorable conditions for wave propagation created by dust-induced atmospheric inflation \cite{liu2019seasonal,liu2023surface}. 
 
\section{NGIMS-PCM Comparisons}
\subsection{Global Circulation of Mars' Upper Atmosphere}
The Martian upper-atmosphere winds exhibit high temporal variability, making it challenging to produce simulations consistent with observations. A common compromise is to compare the averaged wind components (U and V) over specific periods, while disregarding wind orientations. However, this approach overlooks the variability of the wind vector, particularly when the meridional components are significant.

\begin{figure}
\noindent\includegraphics[width=14cm, trim={0cm 0.1cm 0.2 0.1cm}]{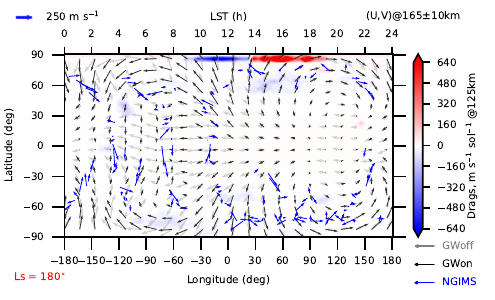}
\caption{Upper winds at 165 $\pm$ 10 km, Ls =180$^\circ$, MY35. Simulations without (gray) and with (black) GWs are hown to illustrate the non-orographic GW's effect. NGIMS-derived winds (blue vectors) are aligned with Local Solar Time (LST). Wave-induced drags are color-coded in the background at an altitude of 125 km.}
\label{glbcclt}
\end{figure}

\citeA{benna2019global} introduced a method to generate “equivalent” simulations that are consistent with the instantaneous winds observed by NGIMS. Their approach takes advantage of the strong diurnal rotation of thermospheric winds predicted by simulations. The observed winds are first aligned with Local Solar Time (LST) and longitude. The aligned winds are then compared with the simulations at Ls = 180$^\circ$ and at an altitude of 150 km. This method ensures that the comparison captures both the climatological averages (the diurnal cycle) and the temporal variability.

Our simulations (Figure \ref{glbcclt}) also predict a strong diurnal rotation of thermospheric winds, both with (black vectors) and without (gray vectors) GWs. The observed winds at 165$\pm$10 km are aligned with LST ( within $\pm$3 h) and longitude during MY35 at Ls 180$^\circ$. The upper atmosphere is in a quasi-isothermal state \cite{gonzalez2010thermal}, which implies highly synchronous wind magnitudes and orientations across altitudes. The simulation of GW divergence at 125 km is color-coded to identify the GW's activities. Most of the waves have saturated below 130 km during this period. The discrepancies (except for some outliers) between the GWon simulations and the observations are within 100 m s$^{-1}$ and likely reflect temporal mismatches and local time offsets.

In the sense of climatology, the observed winds display a diurnal rotation that aligns well with the simulations in terms of magnitude and orientation. Wind vectors propagate from east to west due to enhanced heating during sunrise (LST 6–9 h), producing a counterclockwise rotation in the Northern Hemisphere (NH) and a clockwise rotation in the Southern Hemisphere (SH). 
Because upper-atmospheric winds are largely thermal-wind-driven \cite{gonzalez2010thermal}, they inherit the diurnal thermal tide \cite{forbes2020solar, yang2024martian}. As a result, within LST 3-6 h, the wind perturbations are significantly enhanced due to the constructive superposition of multiple tidal components \cite{forbes2004tides,forbes2020solar}. In particular, the dominant migrating diurnal tide combines with non-migrating tidal modes of different zonal wavenumbers, leading to phase alignment in this local-time sector. This multi-mode interference reinforces the wind variability, producing persistently large amplitudes (from $<100~\mathrm{m\,s^{-1}}$ to $>400~\mathrm{m\,s^{-1}}$ in the observations), consistent with the tidal decomposition framework described in \citeA{forbes2004tides} and \citeA{forbes2020solar}. The waves further modulate the amplitude of this tide-dominated variability (Figure \ref{glbcclt}). Because thermal tides are a primary source of large-scale variability in the Martian middle and upper atmosphere and can both accelerate and damp the zonal mean flow, the observed local-time enhancement is likely embedded within the background tidal phase evolution \cite{forbes2004tides,forbes2020solar, yang2024martian}.

The winds during LST 0-3 h maintain the pattern of the previous sol, in which the vectors experience clockwise rotation in the NH and counterclockwise rotation in the SH. From morning to early afternoon (LST 9–13 h), the winds retain the early-morning diurnal pattern across both hemispheres. The afternoon to sunset winds (LST 13-18 h) rotate clockwise in the NH and counterclockwise in the SH. The night winds (LST 18-24 h) maintain the afternoon pattern, producing cyclonic structures in both hemispheres. The whole cycle is synchronous with the diurnal thermal tide, implying that the diurnal wind pattern is forced by solar energy and the planet's rotation.

Although wave drags are deposited more intensively in high-latitude regions than in the tropics \cite{asumi2025climatology}, winds in the tropics and subtropics are also significantly affected due to wave saturation at lower altitudes. In the polar regions, wind magnitudes are dominated by the zonal component \cite{benna2019global,liu2023surface}. Wind magnitudes respond rapidly to wave drags due to direct interactions between waves and zonal winds. Consequently, high-latitude regions provide favorable locations to observe the behavior of zonal jets under the influence of GWs.



\begin{figure}
\noindent\includegraphics[width=14cm, trim={0.0cm 0.25cm 0.5 0.5cm}]{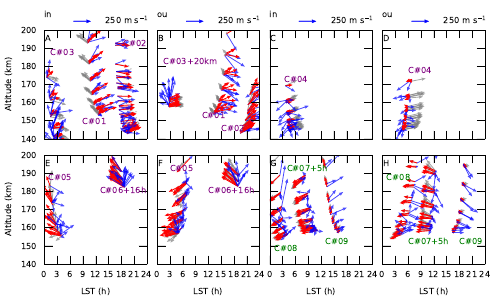}
\caption{Comparisons between PCM simulations and NGIMS winds for \texttt{C\#}01 to \texttt{C\#}09. Gray and red vectors are simulations without and with GWs, respectively. NGIMS observations are in blue. The simulations and observations are sampled at the same longitudes, latitudes, Ls, LST, and altitudes. The data in panel B are shifted upward by 20 km in altitude for better visibility.}
\label{pdf3236}
\end{figure}

\subsection{Waves' Effects on Wind Orientations and magnitudes} 
Figure \ref{pdf3236} presents direct comparisons between simulated and observed winds for the nine cases shown in Figure \ref{obsgeo}. Note that the model provides instantaneous wind fields at 2-hour intervals. If the 1-Hz observations, collected within a 10-minute window (inbound and outbound), fall within a given 2-hour interval, the comparison is performed between the instantaneous observations and the temporally interpolated simulated winds. \texttt{C\#}01-\texttt{C\#}06 are suggested to be strongly influenced by waves (Figure \ref{obsgeo} and \ref{simujets}). The quantitative comparison between simulations and observations is summarized in Table 2.

\begin{table}
\centering
\caption{Vector correlation [0,1] and bulk phase difference (deg) between simulated and observed winds in Figure \ref{pdf3236}. Each entry is formatted as $|r_v|$; $\delta\phi$.}
\label{vector_corr}
\begin{tabular}{l|ccccccccc}
\hline
 Cases& \texttt{C\#}01 & \texttt{C\#}02 & \texttt{C\#}03 & \texttt{C\#}04 & \texttt{C\#}05 & \texttt{C\#}06 & \texttt{C\#}07 & \texttt{C\#}08 & \texttt{C\#}09 \\
\hline
\multicolumn{10}{l}{\textbf{Inbound}} \\
GWon-Obs   & 0.79; 9   & 0.83; -3  & 0.72; -56 & 0.37; 76  & 0.63; 101 & 0.71; 39  & 0.53; -18 & 0.06; -10 & 0.88; 53 \\
GWoff-Obs  & 0.82; 102 & 0.84; 2   & 0.80; -80 & 0.42; 59  & 0.64; 91  & 0.71; 35  & 0.53; -16 & 0.26; 140 & 0.89; 52 \\
\hline
\multicolumn{10}{l}{\textbf{Outbound}} \\
GWon-Obs   & 0.84; 40  & 0.82; 19  & 0.84; -80 & 0.18; -114 & 0.39; 87 & 0.85; 49  & 0.48; -90 & 0.55; 168 & 0.95; 20 \\
GWoff-Obs  & 0.76; 126 & 0.78; 30  & 0.81; -89 & 0.47; 124  & 0.37; 100 & 0.84; 39 & 0.40; -109 & 0.55; 162 & 0.97; 37 \\
\hline
\end{tabular}
\end{table}

Comparisons for \texttt{C\#}01 show that GWs can substantially change the orientations and magnitudes of the winds (Figure \ref{pdf3236}a and \ref{pdf3236}b). Waves can modify both the direction and magnitude of wind vectors. The simulation of wind magnitudes including wave effects (GWon) agrees well with the observations (Both magnitudes are $\sim$ 300 m s$^{-1}$ with a difference $\sim$ 20 m s$^{-1}$; also Table \ref{vector_corr}). The wave's effect is so distinct that it opens a new jet at northern high-latitudes during Ls 270$^\circ$-300$^\circ$ (at the end of the A-storm, the annual dust peak), as shown in Figure \ref{simujets}j. In this case, even instantaneous winds appear to be governed by non-orographic GWs rather than by solar forcing.

\texttt{C\#}02 and \texttt{C\#}03 were observed in the southern polar region during Ls 143$^\circ$ and 125$^\circ$ (Figure \ref{obsgeo}), MY34, respectively. They are located inside the corresponding damped jet in Figure \ref{simujets}e. Weak wave influences affect the winds during the night in both the ingress and outbound legs (Figure \ref{pdf3236}b and \ref{pdf3236}b). The exception is the simulated inbound winds in \texttt{C\#}03, where the GWon case ($\sim$ 100 m s$^{-1}$) shows orientations closer to the observations ($\sim$ 100-320 m s$^{-1}$) than GWoff ($\sim$ 100-250 m s$^{-1}$) . The predawn winds are mainly governed by the diurnal cycle, reflecting the weak wave activity at that LST (Figure \ref{glbcclt}).

Although the simulated wind magnitudes ($\sim$ 80 m s$^{-1}$) in \texttt{C\#}04 are smaller than those observed ($\sim$ 100-300 m s$^{-1}$, Figures \ref{pdf3236}c and d), the wind directions (GWon) generally lie with the measurements. For the outbound case (Figure \ref{pdf3236}d), part of the winds in GWon ($\sim$ 50 m s$^{-1}$) are closely aligned with the observations ($\sim$ 50-200 m s$^{-1}$; Table \ref{vector_corr}). \texttt{C\#}04 is located in the southern polar jet at the end of month 1 (Figure \ref{simujets}a) or at the start of month 2 (Figure \ref{simujets}b).

The simulated wind magnitudes ($\sim$ 250 - 300 m s$^{-1}$) in \texttt{C\#}05 and \texttt{C\#}06 are consistent with the observations (Figure \ref{pdf3236}e and \ref{pdf3236}f). The high variability of wind orientations is not fully captured by the simulations. The jets where the two cases are located are slightly impacted by the waves (Figure \ref{simujets}g and \ref{simujets}l).

\texttt{C\#}07 to \texttt{C\#}09  are located in the ascending branch of HC (Figure \ref{obsgeo}). Thus, the wave effects are minor (Figure \ref{pdf3236}g and \ref{pdf3236}h) because the simulations without and with the GWs are similar to each other.
They are consistent with the predictions in Figure \ref{simujets}b, \ref{simujets}d, and \ref{simujets}e. The discrepancies between simulations and observations in these cases indicate that the winds have high temporal variability that is not revealed by the model.

Figure \ref{simujets} indicates that non-orographic GWs can significantly modulate, and in some cases dominate, the diurnal variation of winds in the polar region during dusty seasons. Although non-orographic GWs can modulate the instantaneous behavior of zonal jets in other seasons (Figure \ref{simujets}), the climatological wind patterns are predominantly controlled by the diurnal cycle induced by solar forcing. The inclusion of gravity-wave effects generally improves bulk directional alignment, although the magnitude improvement depends on the region (Table \ref{vector_corr}).

\subsection{Zonal Jets from NGIMS and Mars PCM}

\begin{figure}
\noindent\includegraphics[width=12cm,trim={0.0cm 0.1cm 0 0.1cm}]{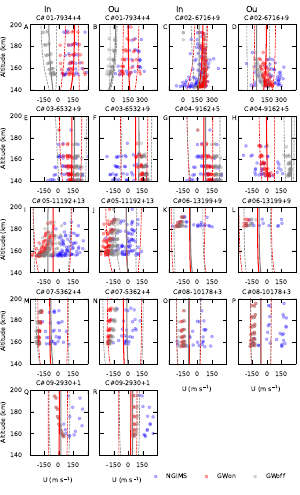}
\caption{Zonal wind magnitudes from NGIMS observations (blue) and PCM simulations (red and gray) for case\texttt{\#}01 to case\texttt{\#}09. Solid lines for model predicted climatology and dash lines for 2-$\sigma$ threshold.
For example, \texttt{C\#}01-7934+4 refers to case 01, starting orbit number 7934 and 4 orbital maneuvers. Note that not all orbit numbers have continuous samples; for instance, \texttt{C\#}05-11192+13 comprises a total of 5 samples.}
\label{pdffigurezonaljets}
\end{figure}

The interactions between wave momentum and mean flows occur predominantly in the zonal direction \cite{lott2012stochastic,lott2013stochastic}. The GW momentum in the meridional direction saturated just above a few layers of the waves' source altitudes due to low background meridional velocity \cite{lott2012stochastic,liu2023surface}. Variations in the meridional winds of the upper atmosphere are driven by zonal drags \cite{liu2023surface, liu2025diurnal, asumi2025climatology}, since the meridional drags approximate zero in the upper atmosphere \cite{liu2023surface,asumi2025climatology}. Here, we focus on high-latitude upper-atmosphere zonal jets.

Figure \ref{pdffigurezonaljets} shows the instantaneous PCM-NGIMS comparisons and the model climatology within $\pm$6 hours of the samples' local solar time. Figures \ref{pdffigurezonaljets}a and \ref{pdffigurezonaljets}b demonstrate that non-orographic GWs contribute to the opening of the zonal jet in the northern hemisphere shown in Figure \ref{simujets}j. The 'opening' effect is evident in both instantaneous and climatological zonal winds (Figure \ref{simujets}j), reflecting the dominant role of non-orographic gravity waves in the diurnal wind cycle during the dusty season.

The "accelerating" is shown in \texttt{C\#}02 (Figure \ref{pdffigurezonaljets}c and \ref{pdffigurezonaljets}d). Most of the observations fall within the ranges of the model climatology, although there are discrepancies ($\sim$ 50-100 m s$^{-1}$) between the instantaneous simulations and the observations. Outliers greater than 300 m s$^{-1}$ are not captured by the simulations. The $\pm$6-hour (accelerating) winds oppose the monthly (damping) climatology shown in Figure \ref{simujets}e, indicating the variability of the winds.

Figures \ref{pdffigurezonaljets}e and \ref{pdffigurezonaljets}f show the damping effect of waves during the clear-sky season. Both the instantaneous and climatological simulations that include non-orographic GWs (GWon) are in closer agreement with the observations than those GWoff cases. Similar damping effects are shown in \texttt{C\#}04. The effect is more pronounced in the outbound leg (Figure \ref{pdffigurezonaljets}h) than in the inbound sample (Figure \ref{pdffigurezonaljets}g). Figure \ref{pdffigurezonaljets}h suggests the closing effect of the waves.

Observations of \texttt{C\#}05 are sampled during the predawn periods. Winds fluctuate in a range of -300  to 300 m s$^{-1}$ from the inbound leg (Figure \ref{pdffigurezonaljets}i) to the outbound leg (\ref{pdffigurezonaljets}j), as a result of interactions with the diurnal cycle (Figure \ref{glbcclt}). The climatology reveals a slight damping effect in the inbound leg, while no discernible effect is evident in the outbound orbit. \texttt{C\#}06 exhibits a limited gravity-wave effect in both legs (Figure \ref{pdffigurezonaljets}k and \ref{pdffigurezonaljets}l), consistent with the monthly climatology shown in Figure \ref{simujets}p. The instantaneous data points partially overlap, whereas a small number of observational outliers are not captured by the simulations. The outliers are also shown in Figures \ref{pdf3236}e and \ref{pdf3236}f. The GWon cases show poorer agreement with the observations than the GWoff cases in \texttt{C\#}05 and \texttt{C\#}06, where the Mars PCM imposes an instantaneous wave-induced acceleration, whereas the observed winds instead require wave damping.

No distinct GW effects are observed in \texttt{C\#}07 to \texttt{C\#}09 (Figure \ref{pdffigurezonaljets}m to \ref{pdffigurezonaljets}r). The jets in the ascending branches of the Hadley Cell are rarely modulated by the GWs during all seasons (Figure \ref{simujets}). The Outliers in \texttt{C\#}08 and \texttt{C\#}09 indicate substantial local wind variability beyond the diurnal cycle (Figure \ref{glbcclt}).

\section{Seasonal Wave-Flow Interactions}
\begin{figure}
\noindent\includegraphics[width=14cm,trim={0.1cm 0.1cm 0.0 0.1cm}]{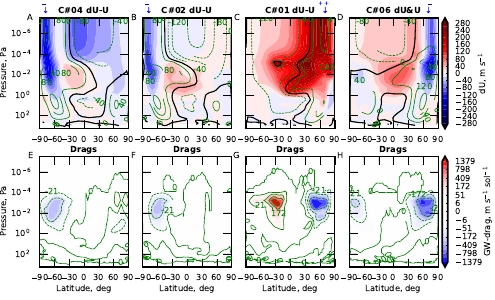}
\caption{Five-sols averaged zonal-averaged PCM simulated fields for cases \texttt{C\#}01, \texttt{C\#}02, \texttt{C\#}04, and \texttt{C\#}06. The first row displays the U-dU (m s$^{-1}$) map; dU is color-coded and U is in contour lines (black lines denote U=0 m s$^{-1}$). The second row depicts the wave's drags (m s$^{-1}$ sol$^{-1}$); Overlapped contour lines to distinct the small values. Note that the contour spacing in the second row is nonlinear.}
\label{pdfpcmfi}
\end{figure}

Four cases are selected to describe the wave-flow interactions during different seasons (Figure \ref{pdfpcmfi}). The simulated zonal winds (U), wave effects (dU), and wave drags are averaged within 5 sols around the observed Ls. All cases are in the descending branches of the HC (see Figure \ref{obsgeo}; Figure \ref{simujets}).

\texttt{C\#}04 (Figures \ref{pdfpcmfi}a), \texttt{C\#}02 (Figure \ref{pdfpcmfi}b), and \texttt{C\#}06 (Figure \ref{pdfpcmfi}d) are near the equinox seasons, during which the HC has two downward branches (Figure \ref{simujets}). In all three cases, the jets are damped by the waves and remain confined to the hemisphere with stronger eastward winds in the middle atmosphere. The eastward winds are stronger in one hemisphere than in the other as a result of the asymmetric of the HC. This favors the propagation of the GWs due to the absence of wind filtering suggested by section 2.1. Thus, a large wave momentum can be deposited in the thermosphere in the corresponding hemisphere (Figure \ref{pdfpcmfi}e, \ref{pdfpcmfi}f, and \ref{pdfpcmfi}h). The jets are thus impacted by wave drag due to wave-flow interactions shown in equation (\ref{divergens}).  The damping effect of \texttt{C\#}02 is the weakest among the three cases due to the weak wave drag (Figure \ref{pdfpcmfi}f).

The upper northern jet in \texttt{C\#}01 (Figure \ref{pdfpcmfi}c) is fully governed by the GW' drags (Figure \ref{pdfpcmfi}g). \texttt{C\#}01 was sampled during the storm peak (Figure \ref{obsgeo}; Figure \ref{simujets}i). Dust storms can create favorable conditions for wave propagation by increasing atmospheric density \cite{liu2019seasonal,han2022thermospheric}, effectively lowering kinematic viscosity ($\nu=\mu/\rho$). Moreover, the northern middle atmospheric eastward winds cause the absence of wind filtering. Similarly, eastward middle-atmosphere winds favor wave growth by preventing their filtering. Thus, the wave momentum is transported to the upper atmosphere instead of breaking at low altitudes (Figure \ref{pdfpcmfi}g). The evolution of wave momentum in different seasons is compatible with the analysis of \citeA{asumi2025climatology}.

Wave–flow interactions are mainly governed by the ’ intrinsic frequency of the waves with Doppler shift ($\Omega$), as described in section 2.1. The horizontal phase speed ($c$) of non-orographic GWs approximates the background zonal wind velocity (U) at the waves' launch altitude \cite{lott2012stochastic, liu2023surface}. In the hemisphere corresponding to the descending branch, eastward winds in the middle atmosphere (Figures \ref{pdfpcmfi}a–\ref{pdfpcmfi}d) exceed the phase speed $c$ in the lower atmosphere, where the waves originate (Figure \ref{simujets}). 

Consequently, the wave’s $\Omega$ is always negative ($k>0$) in the relevant middle-atmosphere region. 
Consequently, $\Omega$ does not change signs between two adjacent atmospheric layers, resulting in the absence of wind filtering. Thus, under these conditions, the middle atmosphere remains free of critical layers. Waves originating in the lower atmosphere traverse the middle atmosphere without momentum deposition and propagate directly into the upper atmosphere (Figure \ref{pdfpcmfi}e to \ref{pdfpcmfi}h). The hemispheric asymmetry of the zonal winds (Figure \ref{simujets}) leads to an asymmetric evolution of momentum flux and drag  (Figures \ref{pdfpcmfi}e-h).

In the upper atmospheric polar region, where the descending branch of the Hadley circulation occurs (Figure \ref{simujets}; Figure \ref{pdfpcmfi}a to \ref{pdfpcmfi}d), the background density ($\rho$) is sufficiently low that the wave amplitudes increase and dissipative processes become more effective. In particular, the resulting increase in kinematic viscosity ($\propto 1/\rho$), together with wave-induced turbulence \cite{lott2012stochastic}, leads to enhanced wave dissipation and the associated transfer of momentum to the mean zonal flow (Figure \ref{pdfpcmfi}e-h).

\section{Summary}
We examine the Martian polar upper-atmospheric zonal jets using simulations from the Mars Planetary Climate Model (Mars PCM) along with observations from the Neutral Gas and Ion Mass Spectrometer (NGIMS). These jets are modulated (damped, closed, accelerated, or even generated) by the momentum released from the non-orographic gravity waves (GWs). 

The jets most strongly influenced by non-orographic GWs are predominantly located in the hemisphere corresponding to the descending branch of the Martian Hadley Cell (HC). This is confirmed by the NGIMS-derived winds.
Strong eastward winds in the middle atmosphere facilitate the propagation of waves within the branches by eliminating wave critical layers \cite{liu2025diurnal,kuroda2016global}. 
Consequently, affected jets exhibit hemispheric asymmetry, reflecting the asymmetry of the Martian HC during seasonal evolution \cite{forget1999improved}.

The jets can be generated or demolished instantaneously by non-orographic GWs due to the momentum released by the waves. The opening and closing captured by the Mars PCM \cite{liu2023surface,liu2025diurnal,liu2025stochastic} can be found in the NGIMS observations.

Wave effects on jets intensify during dusty seasons relative to clear-sky periods, driven by dust-enhanced saturation altitudes and elevated momentum magnitudes \cite{kuroda2020gravity,liu2019seasonal,liu2023surface, liu2025stochastic,asumi2025climatology}. Both simulations and observations show the dominant role of non-orographic GWs in the investigated jets during the MY34 storm peak.

The high variability of the winds in the observations ($>$350-400 m s$^{-1}$) is not fully captured by the simulations during the clear-sky seasons, indicating the existence of undiscovered local dynamics such as thermal tide forcing \cite{forbes2004tides,forbes2020solar,yang2024martian}.

\section*{Open Research}
The Neutral Gas and Ion Mass Spectrometer-derived winds \cite{elrod2014maven} used in this paper are available in the NASA Planetary Data System \url{https://atmos.nmsu.edu/PDS/data/PDS4/MAVEN/ngims_bundle/l3/}. The Mars Planetary Climate Model is available at \url{ svn.lmd.jussieu.fr/Planeto/trunk} (r3263). The simulations can be repeated by running the model with built-in configuration files. The data used in this paper are available at \citeA{liu2026_data}. 

\section*{Comprehensive Conflict of Interest}
None.
\acknowledgments
J. Liu acknowledges the funding from the National Natural Science Foundation of China (Grant 42241115). We gratefully acknowledge the use of HPC computing resources from GENCI-CINES (Grant A0160110391). We are very grateful to François Lott for his clear explanation, which helped us properly distinguish between the parameterized momentum fluxes and the resolved EP-flux formulation. The authors sincerely thank the two anonymous reviewers for their detailed, thoughtful, and constructive comments, which greatly improved the clarity and overall quality of this manuscript.


%
%



\bibliography{agusample}

%
%
%
%
%

\end{document}